\documentclass[runningheads]{llncs}

\usepackage[T1]{fontenc}
\usepackage{graphicx}
\usepackage{amsmath}

\begin{document}
\title{DexAssist: A Voice-Enabled Dual-LLM Framework for Accessible Web Navigation}
%

%
\titlerunning{DexAssist: A Dual-LLM Framework for Accessible Web Navigation}
%
\author{Shridhar Mehendale\inst{1}\orcidID{0009-0000-1539-6616}\textsuperscript{*}\and
Ankit Walishetti\inst{1}\orcidID{0009-0009-1792-4959}\textsuperscript{*}}
%

\authorrunning{S. Mehendale and A. Walishetti}

%
\institute{Illinois Mathematics and Science Academy, Aurora IL 60506, USA}

\renewcommand{\thefootnote}{*}
\footnotetext[1]{Both authors contributed equally to this work.}

\maketitle              
\vspace{-8mm}
\begin{abstract}
Individuals with fine motor impairments, such as those caused by conditions like Parkinson's disease, cerebral palsy, or dyspraxia, face significant challenges in interacting with traditional computer interfaces. Historically, scripted automation has offered some assistance, but these solutions are often too rigid and task-specific, failing to adapt to the diverse needs of users. The advent of Large Language Models (LLMs) promised a more flexible approach, capable of interpreting natural language commands to navigate complex user interfaces. However, current LLMs often misinterpret user intent and have no fallback measures when user instructions do not directly align with the specific wording used in the Document Object Model (DOM). This research presents Dexterity Assist (DexAssist), a dual-LLM system designed to improve the reliability of automated user interface control. Both LLMs work iteratively to ensure successful task execution: the Navigator LLM generates actions based on user input, while the Support LLM assesses the success of these actions and provides continuous feedback based on the DOM content. Our framework displays an increase of $\sim$36 percentage points in overall accuracy within the first iteration of the Support LLM, highlighting its effectiveness in resolving errors in real-time. The main contributions of this paper are the design of a novel dual LLM-based accessibility system, its implementation, and its initial evaluation using 3 e-commerce websites. We conclude by underscoring the potential to build on this framework by optimizing computation time and fine-tuning.

\end{abstract}
\vspace{-8mm}
\section{Introduction}
 \vspace{-3mm}

E-commerce has become an essential component of everyday life for many, with digital platforms like Amazon emerging as key players in the industry. However, as e-commerce websites have evolved, their layouts have become increasingly complex and densely packed with information \cite{Tsagkias2020}. This has caused significant challenges for individuals with conditions such as Parkinson's disease, cerebral palsy, and dyspraxia who lack the fine motor skills to interact with these intricate web interfaces \cite{Muthu2023AssistiveTechnology}. These accessibility challenges have motivated the development of assistive technology reliant on state-of-the-art (SOTA) large language models (LLMs) to autonomously control computer interfaces.

\subsubsection{Previous Works.}
Nogueira and Cho \cite{webnav} developed WebNav, a tool that constructs a website-flow graph that represents each page on the website as a node and links them together to allow the agent to navigate. However, this forces the agent to follow specific node chains, significantly reducing its ability to scale to abstract web tasks.

Soon after, the World of Bits (WoB) benchmark \cite{WofB} allowed RL agents to complete tasks on web pages using DOM analysis. These agents were trained to follow action sequences of humans performing the same task. To broaden decision-making capabilities across a more extensive range of actions and webpages, the authors of \cite{WebShop} developed WebShop. This tool enabled agents to navigate both Amazon and a custom website designed specifically for the research. However, the training accuracy of the dataset itself was relatively low (59\%).


Innovating on such approaches, Y Qin et al. \cite{toollearningfoundationmodels} leveraged SOTA GPT-3.5 to interact with specific websites. The researchers developed web navigation scripts in Python and prompted GPT to invoke the scripts based on the user's requests. The prompts used were either zero-shot (only context about the available scripts was provided) or few-shot (script context and examples of valid queries and responses were provided). However, in either case, if the model failed to perform the task correctly, there were no fallback measures in place. Y Qin et al. tested this framework on a variety of web tasks including online shopping (Amazon specifically). 

\vspace{-6mm}
\subsubsection{Objective.}
This study develops a reliable framework to handle errors when autonomously controlling the web. We develop Dexterity Assist (DexAssist) as a scalable tool for implementing LLMs into web automation tasks, specifically in the context of serving as assistive technology to aid individuals with fine motor impairments.\footnote{Demo videos: \url{https://github.com/smeh25/DexAssist/issues/1}}


\vspace{-5mm}
\section{Methods}
\vspace{-4mm}
This section develops DexAssist, a dual-LLM\footnote{LLama 3.1 8B, the only LLM used, was run locally via Ollama on a Macbook Pro M3 11-core with 18GB Memory and 512 SSD Storage} system to increase the capabilities of automated web scripts, enhancing accessibility for individuals with fine motor impairments who struggle to use traditional computer interfaces. A Navigator LLM is initially provided with information about a set of flexible automation scripts, a list of key DOM elements, and the user's instructions. Then a Support LLM is introduced to monitor the success of the Navigator LLM and provide real-time, iterative feedback.


\vspace{-5mm}
\subsection{Dynamic Automation Scripts}
 \vspace{-3mm}
\label{sec:primary_functions}
To enable flexible and reliable automation of web interactions, we develop a series of Python scripts utilizing the Playwright library, which is chosen for its robustness in handling modern web applications and its ability to effectively control Chrome version 127.0.6533.17. These scripts provide a set of generalized functions that can be easily adapted to a wide range of websites. In Table \ref{functionExamples} we describe the three primary functions used to automate web interactions. Each function contains built-in checks to ensure the successful execution of actions. In cases where an action cannot be completed, comprehensive error codes are returned.

\vspace{-2mm}
\begin{table}[h]
\centering
\vspace{-3mm}
\caption{Primary functions used for automating web interactions.}
\begin{tabular}{|l|p{9cm}|}
\hline
\textbf{Function Header} & \textbf{Description} \\ \hline
\texttt{click(visibleText)} & Locates and clicks on the element matching the provided parameter. \\ \hline
\texttt{searchBar()} & Identifies and selects the main search bar on a page. \\ \hline
\texttt{type(text)} & Types the provided text into the currently active element on the webpage. \\ \hline
\texttt{press(keyName)} & Presses any key on the keyboard, including but not limited to control and navigation keys such as Enter, Tab, and ESC. \\ \hline
\end{tabular}

\label{functionExamples}
\end{table}
\vspace{-12mm}

\subsection{Handling DOM Structure}
\label{sec:dom_structure}
In this section, we develop a web scraper using Beautiful Soup to extract the DOM structure from a webpage. While this provides a comprehensive view of the page, the resulting data often exceeds the input token limit of the Llama 3.1 8B model due to its length. To address this, we employ Llama 3.1 8B to condense the DOM structure by extracting only the key clickable and interactable elements such as product names, dropdown menus, and navigation links.

\vspace{-3mm}
\subsection{Initial Navigator LLM Call}
\label{sec:main_LLM}
We first use the \texttt{speech\_recognition} library in Python to convert the user's spoken instructions into text. This text is then incorporated into a carefully structured prompt that contains the general instructions, the key clickable DOM elements (see Section~\ref{sec:dom_structure}), and information about the available functions and their specific input variables (see Section~\ref{sec:primary_functions}). Finally, the model is instructed to structure its output in a list format with function names followed by arguments separated by double slashes (\texttt{//}) in sequential order.
 \vspace{-2mm}

\begin{quote}
\textbf{Example of Basic User Instruction and Model Output:}

\textbf{User Instruction:} \textit{I am looking for a pair of shoes under \$50}

\textbf{Model Output:} \texttt{searchBar 0\footnote{The "0" indicates no arguments should be passed} // type shoes under \$50 // press enter}

\end{quote}






 \vspace{-4mm}



\subsection{Dual LLM Structure}
\begin{figure}[h]
\centering
\includegraphics[width=0.6\textwidth]{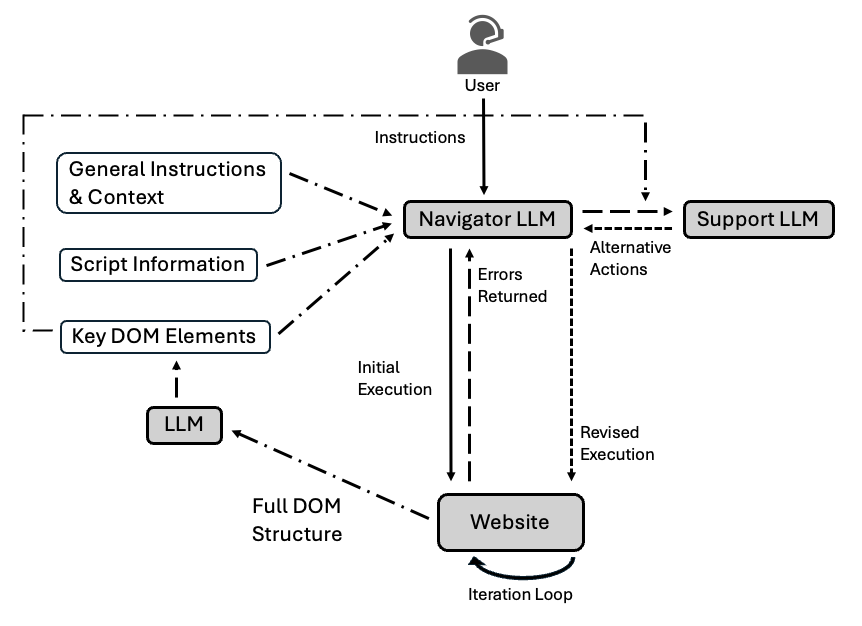}
\caption{Visual Diagram of the Dual-LLM DexAssist Framework
} \label{Framework_Diagram}
\end{figure}

The dual-LLM system handles complex user interactions with web interfaces by employing two distinct language models: the Navigator LLM $\mathcal{N}$ and the Support LLM $\mathcal{S}$. This section details the iterative process that occurs when the Navigator LLM $\mathcal{N}$ encounters execution errors and the Support LLM $\mathcal{S}$ engages to resolve them. The process is described in the following steps and illustrated in Fig. \ref{Framework_Diagram}:

\begin{enumerate}
    \item \textbf{User Instruction:} The user provides an instruction $U_i$ from the set of 100 distinct instructions ${i \in \{1, 2, \dots, 100}\}$, which is processed by the Navigator LLM $\mathcal{N}$ to generate a set of actions $\mathbf{A} = \{a_1, a_2, \dots, a_n\}$.
    \item \textbf{Execution and Error Detection:} Each action $a_i \in \mathbf{A}$ is executed sequentially, and the success of each action $a_i$ is determined by evaluating the execution result $\mathcal{E}(a_i)$. Specifically, $\mathcal{E}(a_i) = 0$ indicates success, while $\mathcal{E}(a_i) \neq 0$ indicates failure. The failed actions are grouped into an error set $\mathbf{E} = \{\mathcal{E}(a_1), \mathcal{E}(a_2), \dots, \mathcal{E}(a_p)\}$. 
    \item \textbf{Support LLM Invocation:} The Support LLM $\mathcal{S}$ is invoked with the set of error codes $\mathbf{E}$ and the DOM structure $\mathcal{D}$. It returns a set of action subsets $\mathbf{A}' = \{\mathbf{A}'_1, \mathbf{A}'_2, \dots, \mathbf{A}'_g\}$, where each subset $\mathbf{A}'_j$ contains a list of possible actions corresponding to each error code $\mathcal{E}(a_i) \in \mathbf{E}$, where $i$ represents any index within the set.
    \item \textbf{Revised Action Execution:} The Navigator LLM $\mathcal{N}$ selects one action from each action subset $\mathbf{A}'_j$ in $\mathbf{A}'$ and attempts to execute these new actions.
    \item \textbf{Iteration and Termination:} The process repeats up to a maximum of $k=4$ iterations. If the issue is not resolved within $k$ iterations, the system terminates and returns a final failure explanation to the user.
\end{enumerate}



The iterative sequence can be summarized as follows: \(\mathbf{A}_{\text{new}} = \mathcal{N}\left(\mathcal{S}(\mathbf{E}, \mathcal{D})\right)\).

\vspace{-6mm}

\vspace{-10mm}
\subsection{Modeling Framework}

To evaluate our framework, we test 100 distinct user instructions $\{U_i \mid i \in \{1, 2, \dots, 100\}\}$ on three major e-commerce websites: Amazon, Target, and Walmart. We measure the cumulative accuracy and compute time achieved across iterations of the dual-LLMs, performing $m=20$ randomized trials to calculate an average. Within each trial, 30 instructions are randomly selected from the set of 100 user instructions. Such e-commerce websites are selected due to their critical relevance in everyday life and their reliance on complex user interactions.

 \vspace{-5mm}
\section{Results}
\vspace{-3mm}

\begin{figure}
\centering
\includegraphics[width=0.8\textwidth]{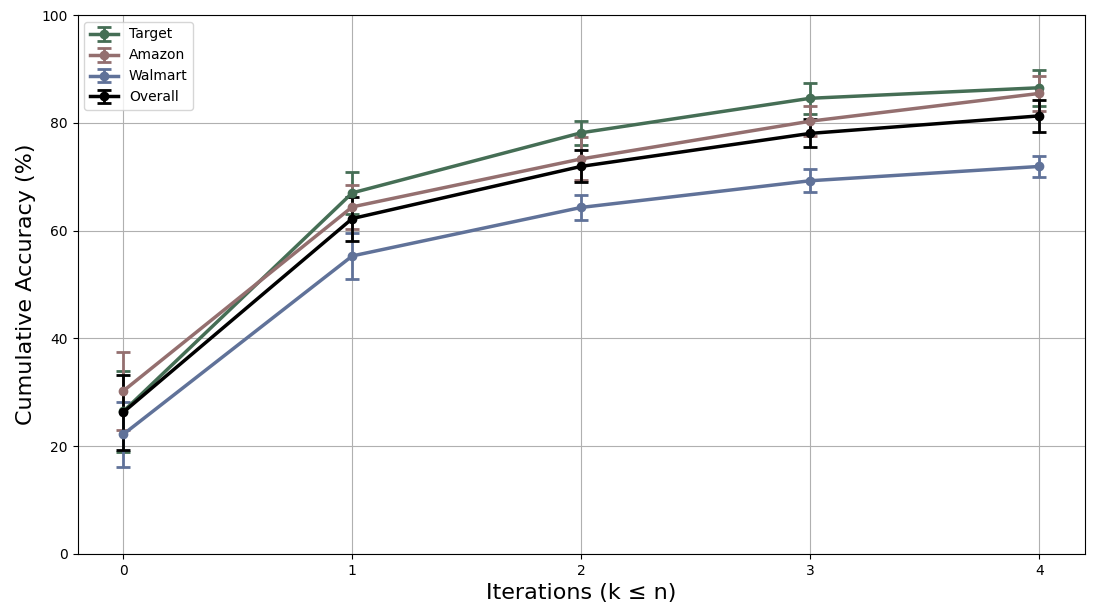}
\caption{Cumulative accuracy averaged over $m = 20$ trials of 30 randomly selected distinct user instructions $U_i$ from $\{U_i \mid i \in \{1, 2, \dots, 100\}\}$ across 3 e-commerce websites. The accuracy is calculated at each iteration $k \leq 4$ and averaged across all trials. Error bars represent the standard deviation $\sigma_k$ for each $k$ value.
} \label{accuracy_vs_iterations}
\end{figure}


In Fig. \ref{accuracy_vs_iterations}, we tested for the cumulative accuracy across \emph{k} iterations across three websites. The results indicate a low initial overall accuracy at $k = 0$ ($\sim$26\%), reflecting the inherent difficulty of the tasks without iterative refinement. Notably, between $k = 0$ and $k = 1$, there is a significant improvement in overall accuracy ($\sim$36 percentage points), supporting the iterative feedback mechanism between the Navigator LLM $\mathcal{N}$ and Support LLM $\mathcal{S}$ is effectively resolving errors and refining the system's actions. The overall cumulative accuracy $\text{Acc}(k)$ converges after $k = 2$, toward an upper bound such that $\lim_{k \to 4} \text{Acc}(k) \approx \text{Acc}_{\text{max}}$, indicating efficiency within the framework.

 \vspace{-6mm}
\begin{table}[h]
\centering
\caption{Average Time for Iterations Across Different Websites (in seconds)}\label{time_iteration_table}
\begin{tabular}{|l|c|c|c|c|c|}
\hline
Website & $k \leq 0$ & $k \leq 1$ & $k \leq 2$ & $k \leq 3$ & $k \leq 4$ \\
\hline
Amazon & 4.54 & 8.67 & 11.73& 13.67 & 14.70\\
Target & 4.24 & 7.89 & 9.34 & 12.32 & 14.47 \\
Walmart & 7.06 & 12.95 & 16.88 & 19.35 & 21.22\\
\hline
\end{tabular}
\end{table}
\vspace{-6mm}

The cumulative average time for iteration \( k \leq n \) for three e-commerce websites: Amazon, Target, and Walmart is reported in Table \ref{time_iteration_table}. Results are averaged over $m=20$ trials with randomized user instructions. The total time taken by the system to complete a task after \( k \leq n \) iterations is represented by the sum of the times for all previous iterations, expressed mathematically as \(\sum_{k=0}^{n} t_k\), where \( t_k \) represents the time for each individual iteration \( k \). As \( k \to 4 \), the change in time between each iteration decreases as there are fewer errors in the output returned by the Navigator LLM \(\mathcal{N}\) as the iterations progress, \( |\mathbf{E}_{k+1}| < |\mathbf{E}_k| \). The results indicate that the Support LLM \(\mathcal{S}\) resolves a portion of these errors in each iteration. For unresolved errors, the system has already explored multiple solutions, enabling it to refine its approach and deepen its understanding, thereby minimizing the time required for subsequent iterations. Furthermore, the results indicate that Walmart's website consistently requires more time per iteration, with a cumulative time of 21.22 seconds at $k \leq 4$, compared Amazon (14.70 seconds) and Target (14.47 seconds). A later manual analysis showed that the Walmart page DOM structure is more complex than Target and Amazon's, making it more challenging for the model to navigate efficiently. This finding aligns with Fig. \ref{accuracy_vs_iterations}, where Walmart had the lowest cumulative accuracy.



 \vspace{-4mm}
\section{Conclusion and Future Work}
 \vspace{-2mm}
Individuals with fine motor impairments who face significant challenges using traditional web interfaces lack comprehensive accessibility tools. This research developed DexAssist, a dual-LLM framework to provide a scalable and reliable system to navigate the web through voice commands. DexAssist displayed marked improvements in task completion accuracy during the initial iterations, particularly when the Support LLM was engaged, effectively resolving errors early in the process. However, early iterations within our current approach are time and resource-intensive. Future work should focus on optimizing the iterations to completing as many iterations as possible (ideally $k \geq 3$) within 9 seconds. Furthermore, future work can build on this framework to implement fine-tuned models that possess highly domain-specific knowledge and more powerful and quicker LLMs.

\vspace{-2mm}
\begin{credits}
\subsubsection{\ackname} The authors would like to thank Dr. Steve Oney from Accessibility and Computing at the University of Michigan and Dr. Lawrence Angrave from the Universal Design for Learning and Accessibility Research Group for their support throughout the design, execution, and manuscript writing process.


\end{credits}


%
%
%
  \vspace{-4mm}

\bibliographystyle{splncs04}
\bibliography{references}

\end{document}